\documentclass[useAMS,usenatbib]{mn2e}
\usepackage{epsfig,lscape} 
\usepackage{amsmath,xcolor}
\usepackage{comment}

\title[Age-chemical structure of the Galaxy I]{Age-chemical abundance structure of the {Galaxy I}: Evidence for {a late accretion event} in the outer disc at $z\sim0.6$}

\author[J. Lian et al.]
{{Jianhui Lian}$^{1}$, {Daniel Thomas}$^{1}$, {Claudia Maraston}$^{1}$, {Olga Zamora$^{2,3}$}, {Jamie Tayar$^{4}$}, \newauthor {Kaike Pan}$^{5}$, {Patricia Tissera}$^{6}$, {Jos\'e G. Fern\'andez-Trincado}$^{7}$, 
	\newauthor {Domingo Anibal Garcia-Hernandez$^{2,3}$}  
	\\
	\small $^{1}$ Institute of Cosmology and Gravitation, University of Portsmouth, Burnaby Road, Portsmouth, UK, PO1 3FX\\
	\small $^{2}$Instituto de Astrofısica de Canarias, 38205 La Laguna, Tenerife, Spain \\
	\small $^{3}$Universidad de La Laguna (ULL), Departamento de Astrofísica, E-38206 La Laguna, Tenerife, Spain \\
	\small $^4$Institute for Astronomy, University of Hawaii, 2680 Woodlawn Drive, Honolulu, HI 96822, USA\\
	\small $^{5}$Apache Point Observatory P.O. Box 59, Sunspot, NM 88349\\
	\small $^{6}$Universidad Andres Bello, Departamento  de Ciencias Físicas, 700 Fernandez Concha, Santiago, Chile\\
	\small $^{7}$Instituto de Astronom\'ia y Ciencias Planetarias, Universidad de Atacama, Copayapu 485, Copiap\'o, Chile\\
}

\begin{document}
	\maketitle
	
	\begin{abstract}
		We investigate the age-chemical abundance structure of the outer Galactic disc at a {galactocentric distance} of $r>10$ kpc as recently revealed by the SDSS/APOGEE survey. 
		{Two} sequences are present in the [$\alpha$/Fe]-[Fe/H] plane with systematically different stellar ages. Surprisingly, the young sequence is less metal-rich, suggesting a recent dilution process by additional gas accretion. As the stars with the lowest iron abundance in the younger sequence also show an enhancement in $\alpha$-element abundance, the gas accretion event must have involved a burst of star formation. In order to explain these observations, we construct a  
		chemical evolution model. In this model we include a {relatively} short episode of gas accretion at late times on top of an underlying secular accretion over long timescales. Our model is successful at reproducing the observed distribution of stars in the three dimensional space of [$\alpha$/Fe]-[Fe/H]-Age {in the outer disc}.
		We find that a late-time accretion with a delay of {$8.2\;$Gyr} and {a timescale of 0.7 Gyr} best fits the observed data, in particular the presence of the young, metal-poor sequence. Our best-fit model further implies that the amount of accreted gas in the late-time accretion event needs to be about three times the local gas reservoir in the outer disc at the time of accretion in order to sufficiently dilute the metal abundance.
		{Given this large fraction}, {we interpret the late-time accretion event as a minor merger {presumably} with a gas-rich dwarf galaxy with a mass $M_*<10^{9}\; M_{\odot}$ and a gas fraction of $\sim 75$ per cent.}

	\end{abstract}
	
	
	\section{Introduction}
	Stars serve as a fossil record in galaxies carrying key information about a galaxy's formation history. In particular, the chemical abundances of stars are like footprints of the interstellar material from which the stars were formed. The interstellar medium (ISM) in a galaxy is enriched with chemical elements through the cycle of stars forming, evolving and dying. Therefore, the investigation of the chemical composition of stars and stellar populations is a powerful tool to unfold the chemical enrichment and formation history of a galaxy. 
	The Milky Way (MW) provides an ideal laboratory to study galactic chemical evolution and test galaxy formation and evolution models because of our ability to resolve stellar populations into individual stars and to measure detailed chemical element abundances.
	
	The abundances of iron (Fe) and the $\alpha$ elements (e.g., Mg, O, Ca, ...) are of particular interest and widely used to constrain the chemical evolution of our Galaxy. The distribution of iron abundance, i.e.\ the metallicity distribution function (MDF) has been used to constrain chemical evolution models for a long time (e.g., \citealt{schmidt1963,pagel1975}) {and more recently also in a cosmological framework \citep{tumlinson2010}}. It was shown early on that the observed MDF in the solar neighbourhood disfavours a closed-box model {(\citealt{vandenbergh1962,pagel1975}), as the latter overestimates the density of metal-poor stars ($[{\rm Fe}/{\rm H}]<-0.2$ dex)}. {A leaky/accretion model including gas inflow and/or outflow is one of the preferred solutions to this problem (e.g., \citealt{audouze1976,pagel1997})}. The observed abundance of $\alpha$-elements combined with Fe further constrains the timescale of star formation. This is because $\alpha$-elements are released promptly through SN-II explosions of massive stars {(with a time delay of only a few 10\;Myr)} while the enrichment of Fe is dominated by SN-Ia explosions of long-lived, low-mass stars {(with longer delay times from a few 10\;Myr to a few 10\;Gyr)}. Hence a low ratio of [$\alpha$/Fe] indicates a long period of star formation, while a high [$\alpha$/Fe] suggests that star formation must have {occurred} on short timescales such that {SN-Ia} did not have the time to release iron \citep{matteucci1994}. This powerful diagnostic has been extended to galaxies, demonstrating that the most massive galaxies formed their stars on the shortest timescales due to their enhanced [$\alpha$/Fe] ratios \citep{thomas2010}.
	
	
	One of the most interesting discoveries regarding the chemical compositions of stars in the MW is the bimodal distribution in the [$\alpha$/Fe]-[Fe/H] plane in which most stars lie on two well-separated sequences \citep{fuhrmann1998,reddy2006,adibekyan2012,haywood2013,bensby2014}. One sequence comprises stars with super-solar $\alpha$-abundance ([$\alpha/{\rm Fe}]>0.15$) while the other consists of stars with solar-like abundance ratios ([$\alpha/{\rm Fe}]\sim 0$). The high-$\alpha$ sequence is generally more metal-poor than the low-$\alpha$ sequence, but both of them span a wide range in metallicity. There is a significant overlap in metallicity between the two sequences. This pattern was first identified for the solar neighbourhood and then recently confirmed to be valid for a large portion of the Milky Way disc \citep{hayden2015}. 
	
	
	Besides chemical composition, age is a further fundamental stellar parameter providing deep insight into the Galaxy's formation history. Combining the observed ages and chemical compositions of stars in the MW allows us to unfold the chemical enrichment history in great detail. However, stellar age measurements are challenging. One way to obtain stellar ages is to match well-determined stellar parameters $T_{\rm eff}$, $\log g$, and [Fe/H] to the isochrones of stellar evolution models {(e.g., \citealt{jorgensen2005})}. An alternative method combines asteroseismology and spectroscopic observations to set constraints on the stellar mass and then derive the stellar age {(e.g., \citealt{pinsonneault2014})}. Both these methods require high-quality spectroscopy.
	
	For a long time, spectroscopic observations were limited to the stars in the solar vicinity (e.g., \citealt{haywood2013,bergemann2014}). A number of recent, large spectroscopic surveys such as SDSS/APOGEE \citep{majewski2017}, LAMOST \citep{zhao2012}, GALAH \citep{ desilva2015} have improved the situation significantly, providing high-quality optical/infrared spectra with well-determined element abundances and other stellar properties for $\sim 10^5$ stars across the Galactic disc. With these large spectroscopic surveys, it is now possible to map the age-chemical abundance structure of the Galactic disc well beyond the solar radius. 
	For example, \citet{ness2016} derive the ages of 70,000 stars from high-quality APOGEE spectra. The authors apply a machine learning algorithm to estimate the stellar mass from the high-resolution spectra and to then infer the age. A group of stars with more reliable masses from asteroseismology observations \citep{pinsonneault2014} are used as a training set. A similar approach is used by \citet{wu2019} to derive the masses and ages for stars observed by the LAMOST survey.
	
	
	
	{In this new era of large spectroscopic surveys reaching well beyond the solar neighbourhood, theoretical models {that are capable to explain} the new data are needed to improve our understanding of MW formation and evolution.} We therefore started a campaign in which we use our chemical evolution model \citep{lian2018a} to analyse the large sample of stars with age and chemical composition measurements from the APOGEE survey. Our ultimate goal is to shed light on the physical mechanisms responsible for the bimodality in the [$\alpha$/Fe]-[Fe/H] plane and the formation of the two separate populations in the Galactic disc. Given the complexity of the topic, we present this work in separate papers. In this first paper we focus on the combined age-chemical abundance structure of the {\it outer} Galactic disc at $r>10$ kpc. The {next} paper of this series will focus on the combined age-chemical abundance structure of the {inner disc and the bulge}. 
	
	{In the present paper we propose a new outer disc formation scenario in which a late-time gas accretion event is invoked to explain the coexistence of old metal-rich and young metal-poor stars in the outer disc. A similar idea of delayed gas accretion was proposed recently in chemical evolution models  \citep{noguchi2018,haywood2019,spitoni2019} to explain the [$\alpha$/Fe]-[Fe/H] distribution in the solar neighbourhood. A similar scenario was previously presented in the semi-analytical model by \citet{calura2009}. 
	This scenario gets further support from recent cosmological simulations reporting the existence of a bimodality in the $[\alpha/{\rm Fe}]$-[Fe/H] sequence caused by the occurrence of two phases of disc formation through gas accretion \citep{grand2017,mackereth2018,clarke2019}. Here we use a chemical evolution model to provide further constraints on a possible late accretion phase based on the age-chemical abundance structure of the outer disc. Our analysis goes beyond the solar neighbourhood providing evidence of an accretion event {affecting the disc at radii at least up to two times the solar radius}.} 

	The structure of this paper is organised as follows. We introduce the sample selection and observational results in \textsection2. A general introduction of our chemical evolution model is presented in \textsection3. A {direct} comparison between models and observations with an illustration of our best-fit model is included in \textsection4. We then discuss the implications of the results in \textsection5 and conclude with \textsection6.

	\section{Data} 
	
	The sample analysed here comprises red clump and red giant stars that were observed by the Apache Point Observatory Galactic Evolution Experiment (APOGEE) survey \citep{majewski2017}, which is part of the Sloan Digital Sky Survey \citep[SDSS;][]{blanton2017}. APOGEE targets primarily horizontal branch and red giant branch stars throughout the Milky Way's bulge, disc, and halo \citep{zasowski2013,zasowski2017}, using the 2.5~m Sloan Telescope at Apache Point Observatory \citep{gunn2006}, the 2.5~m du~Pont telescope at Las Campanas Observatory. Both telescopes are equipped with high-resolution, $H$-band spectrographs \citep{wilson2012,wilson2019}.  Data are reduced and stellar velocities, parameters, and abundances are determined using custom pipelines described in \citet{nidever2015} and \citet{garciaperez2016}.  
	Here, we use stellar metallicities, $\alpha$-element abundances, and stellar ages from \citet{ness2016}, which are based on spectra and pipeline parameters released as part of SDSS/APOGEE Data Release 12 \citep[DR12;][]{alam2015,holtzman2015}.
	Stellar coordinates are taken from APOGEE's DR14 catalogue \citep{abolfathi2018,holtzman2018}, and we use the ``NMSU'' stellar distances in the APOGEE DR14 distances Value Added Catalogue\footnote{https://www.sdss.org/dr14/data\_access/value-added-catalogs/?vac\_id=apogee-dr14-based-distance-estimations}. We verified that our results do not depend significantly on the choice of set of stellar distances.  
	
	\subsection{Sample selection}
			\begin{figure}
		\centering
		\includegraphics[width=8cm]{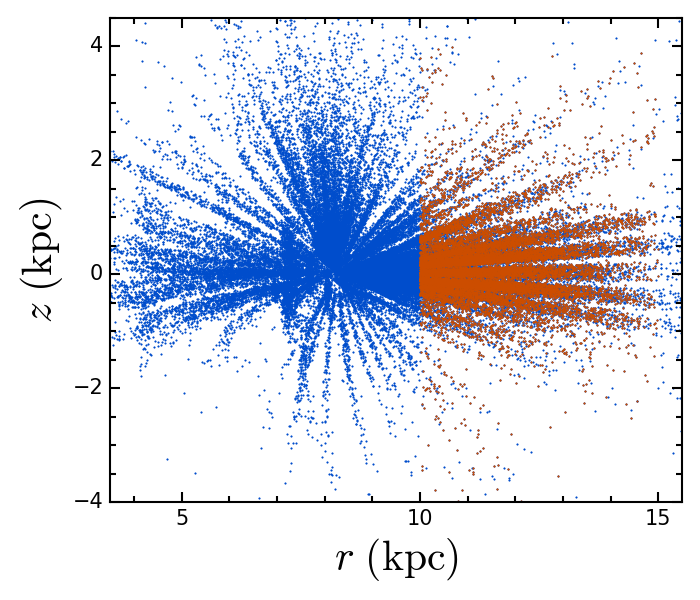}
		\caption{Spatial distribution of stars in the parent sample (blue dots) and our target sample {in the outer disc} with $r>10$ kpc and $S/N>60$ (brown dots). The $x$-axis indicates the {cylindrical} radial Galacto-centric distance, the $y$-axis is the vertical distance from the Galactic mid-plane.}
		\label{spatial}
	\end{figure}
	To ensure a high quality of stellar parameter measurements, we use an additional cut in signal-to-noise (SNR), selecting stars from the parent sample with SNR above 60. We further exclude stars with low-quality spectra as indicated by the flags BAD and WARN in the APOGEE catalogue. We also exclude stars with unreliable parameter estimates selecting stars with flag$_{[\alpha/{\rm Fe}]}=0$ and flag$_{\rm [Fe/H]}=0$ in the star-level bitmask. 
	
	Fig.~\ref{spatial} shows the spatial distribution of the stars in the parent sample (blue dots) in the $r$-$z$ plane. The $x$-axis indicates the {cylindrical} radial Galactocentric distance, the $y$-axis is the vertical distance from the Galactic mid-plane. In the present paper we focus on the {\it outer} disc with $10<r<15$ kpc. The brown dots indicate the spatial distribution of the final selected sample.
	
\begin{figure*}
		\centering
		\includegraphics[width=18cm]{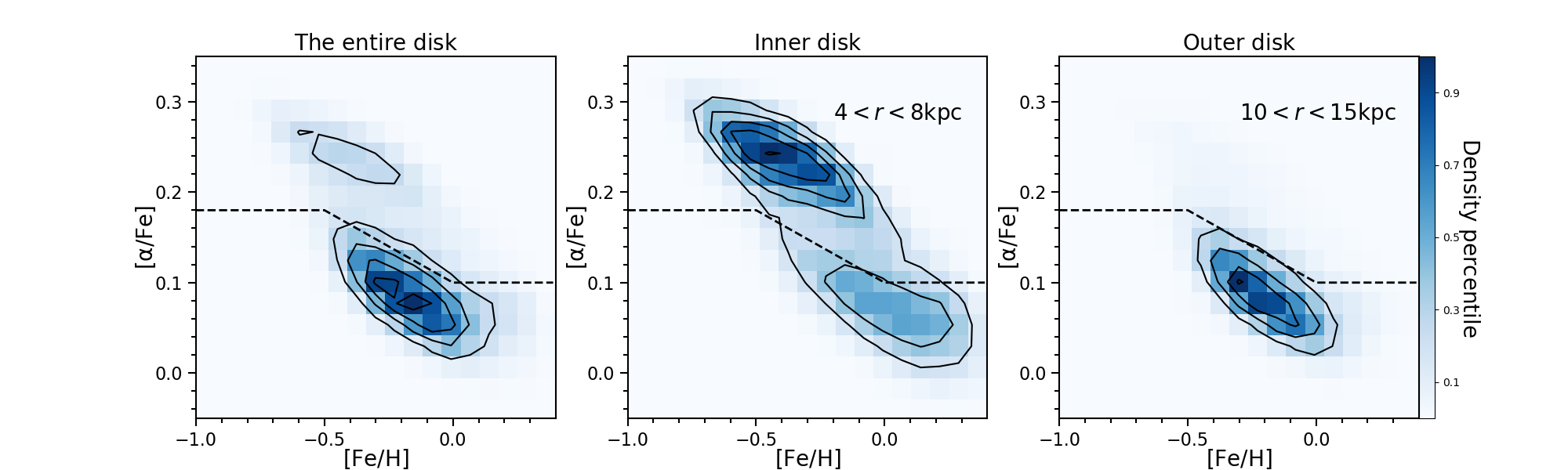}
		\caption{{Density distribution} of stars observed by APOGEE in the [$\alpha$/Fe]-[Fe/H] plane, {with contour lines,} for the entire disc (left-hand panel), the inner disc ($4<r<8$ kpc, middle panel), and the outer disc ($r>10$ kpc, right-hand panel). 
			The bimodal distribution can be clearly seen, and the two populations are well separated. The {dashed} black line highlights the separation.    
			In this work we focus on the Galactic outer disc which is dominated by the low-$\alpha$ population (right-hand panel).}
		\label{feh-afe-inout}
	\end{figure*}

	 We know from observations of the solar neighbourhood that the disc of the Milky Way contains two distinct stellar populations; an old population with high [$\alpha$/Fe] ratios and a young/intermediate-age population with low [$\alpha$/Fe] ratios (e.g. \citealt{fuhrmann1998}). Recent large spectroscopic surveys of the stars in the Milky Way (e.g., \citealt{adibekyan2011,hayden2015}) have revealed that there is a clear bimodal distribution of disc stars in the [$\alpha$/Fe]-[Fe/H] plane as shown in Figure~\ref{feh-afe-inout}. The left-hand panel shows the distribution for the entire disc, while the middle and right-hand panels are for the 
	inner ($4<r<8$ kpc) and outer ($10<r<15$ kpc) discs, respectively. It can be seen that a clear bimodality in the distribution is present {in the inner disc but disappears in the outer disc}.
	
	These two distinct stellar populations are usually referred to as the chemical thick and thin discs. The chemical thick disc is represented by the high-$\alpha$ branch while the chemical thin disc is composed of the low-$\alpha$ branch. To separate the two populations \citet{adibekyan2011} proposed an empirical solution that is based on the minimum number density between the two branches. A similar form of the separation but based on visual inspection {of} our sample is shown as the {dashed} black lines in Figure~\ref{feh-afe-inout}. 
	
	In the inner disc, {a significant fraction of both} low- and high-$\alpha$ populations coexist while the outer disc is {clearly} dominated by the low-$\alpha$ population. 
	This suggests that the formation of the low-$\alpha$ population in the disc is physically decoupled from the high-$\alpha$ population. In this work we focus on the Galactic outer disc ($10<r<15$ kpc) and on the formation of the low-$\alpha$ population. The final sample contains 15,722 disc stars shown in the right-hand panel of Figure~\ref{feh-afe-inout}.
	
	

	
	\subsection{Element abundance and age determination}

	
	The element abundance and age measurements are adopted from \citet{ness2016}. State-of-the-art stellar ages come from asteroseismology observations which provide information about stellar interiors and hence their masses. The stellar mass then sets strong constraints on the stellar age for post-main-sequence stars \citep{martig2015}. The {fractional} uncertainty in the age derived in this way can be as low as $\sim 20$ per cent {(i.e.\ the 1$\sigma$ scatter at the age of 5\;Gyr is 1\;Gyr, for example)}.
	
	Asteroseismology surveys in the literature generally only cover a relatively small portion of the Galaxy \citep{borucki2010}. Moreover, due to the distance limitation in the asteroseismology data, the stellar sample with accurate seismic ages is usually restricted to relatively nearby stars in the solar neighbourhood. The currently largest stellar sample with seismic ages is the APOKASC catalogue \citep{pinsonneault2018} based on Kepler and APOGEE data. The latest version of the APOKASC catalogue contains 5,442 stars. 
	
	Given these limitations, \citet{ness2016} developed a data-driven link between the asteroseismic data for nearby stars and the spectroscopic data for distant stars based on a machine learning algorithm. This calibration allowed them to determine chemical abundances ([Fe/H] and [$\alpha$/Fe]), masses and ages of giant stars directly from spectra. An early version of the APOKASC catalogue \citep{pinsonneault2014} containing 1,639 stars with spectroscopic measurements {($T_{\rm eff}$, [$\alpha$/Fe], [Fe/H]) from APOGEE DR12 \citep{holtzman2015} and seismic measurements (stellar mass, $\log g$) from Kepler was used as a training sample.} Stellar ages were then derived by matching these stellar parameters to PARSEC isochrones \citep{bressan2012}. Base on this training set \citet{ness2016} estimated the ages of 70,000 APOGEE stars from APOGEE spectra. The typical uncertainty of the ages is 40 per cent, hence 0.2 dex. 
	
	\subsubsection{Assessment of the Ness et al. ages}
		\begin{figure}
		\centering
		\includegraphics[width=8cm]{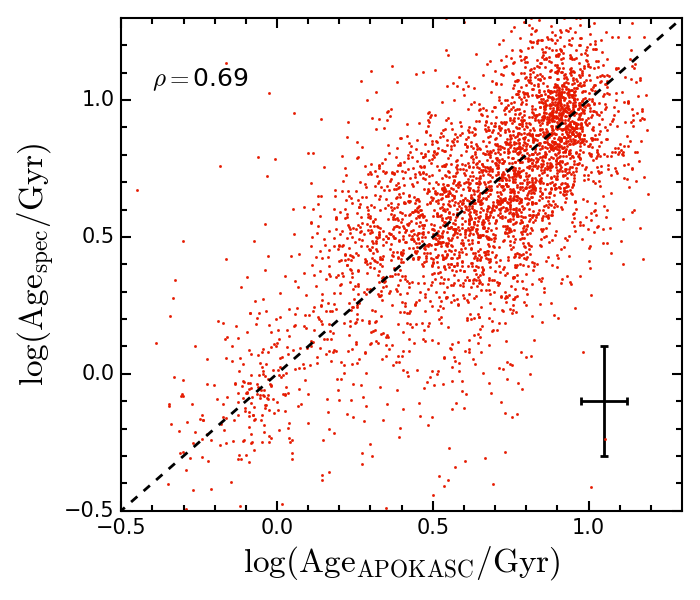}
		\caption{Comparison between the age based on a combination of asteroseismology and APOGEE spectroscopy ($x$-axis) with the age based on APOGEE spectra alone ($y$-axis). Typical uncertainties are indicated by the error bar in the bottom-right corner. The dashed line is the one-to-one relation. There is a good correlation between the two age measurements with a Pearson’s correlation coefficient of 0.69 (top-left corner). The average scatter is 0.26 dex.}
		\label{age-check}
	\end{figure}
	We perform an additional test in order to check the reliability of the spectroscopic ages used here. As already mentioned the ages are calibrated to the seismic ages from the older version of APOKASC with 1,639 stars. With updated Kepler observations, the latest version of the APOKASC catalogue has now been expanded to 5,442 stars. Therefore, we can directly assess the quality of these ages by comparing them with the new seismic ages of those 3,803 stars that were not included in the original training set.
	
	The comparison is shown in Figure~\ref{age-check}. The dashed line indicates the one-to-one relation. It can be seen that the ages from these two measurements are well correlated with a Pearson’s correlation coefficient of 0.69. The scatter of the correlation is 0.26 dex on average. This suggests the ages from Ness et al.\ used in this work are robust within an uncertainty of 0.26 dex.
	
	Note that the scatter in Figure~\ref{age-check} slightly decreases with increasing age. To account for the age-dependent uncertainty we therefore estimate the uncertainties for following three age bins:
	\begin{itemize}
	    \item ${\rm Age}<2.5\;$Gyr ($\log({\rm Age/Gyr})<0.4$)
	    \item $2.5<{\rm Age}<6$ Gyr ($0.4 <\log({\rm Age/Gyr})<0.8$)
	    \item ${\rm Age}>6$ Gyr ($\log({\rm Age/Gyr})>0.8$)
	\end{itemize}
	The adopted scatter is summarised in Table~\ref{tab:agescatter}.
	\begin{table}
		\centering
		\caption{Uncertainties in the age determination.}
		\begin{tabular}{cccc}
			\hline\hline
			Age bin & Scatter & Age bin &  Scatter\\
			(log/dex) & (log/dex) & (linear/Gyr) &  (linear/Gyr)\\
			\hline
			$<0.4$ & $0.3$ & $<2.5$ & $1.8$\\
			$0.4$-$0.8$ & $0.22$ & $2.5$-$6$ & $2.3$\\
			$>0.8$ & $0.19$ & $>6$ & $4.2$\\
			\hline
		\end{tabular}
		\label{tab:agescatter}
	\end{table}
	These errors are mainly caused by the relatively large uncertainties in the spectroscopic age measurements indicated by the error bar in Figure~\ref{age-check}. 
	
		\begin{figure*}
		\centering
		\includegraphics[width=18cm]{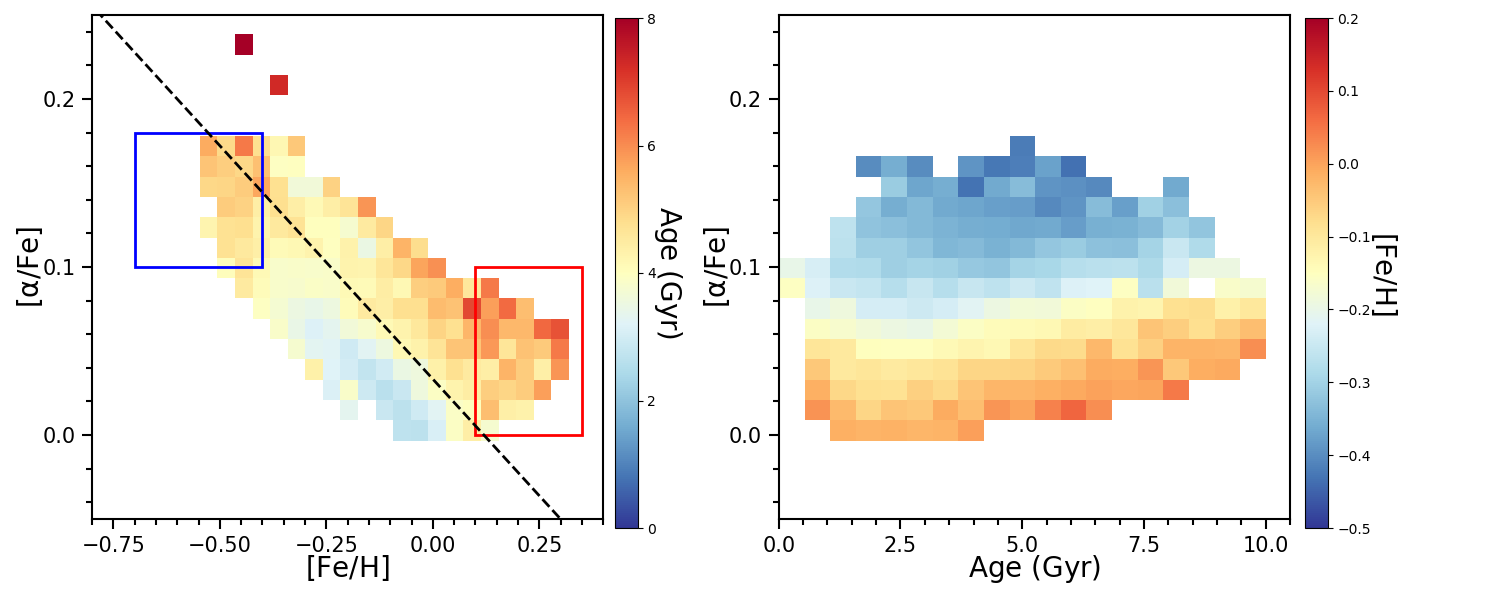}
		\caption{The outer disc stars in the 3-dimensional parameter space of [$\alpha$/Fe], [Fe/H], and Age. {\it Left-hand panel}: [$\alpha$/Fe] as a function of [Fe/H] colour-coded by mean stellar age. The dashed line separates the two sequences with different stellar ages. {The blue and red boxes highlight the location of stars possibly formed at the beginning and at the end point of the late-accretion event, respectively. Their ages {are} used to estimate the start time and the timescale of the {late-accretion event}.}  {\it Right-hand panel}: [$\alpha$/Fe] as a function of age colour-coded by mean [Fe/H]. Two surprising features are present in the plot: 1) an age-metallicity anti-correlation (the presence of young stars with low [Fe/H] as well as old stars with high [Fe/H]), 2) the young metal-poor population is $\alpha$-enhanced (see more detail in the text).}
		\label{afe-feh-age-thin}
	\end{figure*}
	
	
	
	\subsection{Combined [$\alpha$/Fe]-[Fe/H]-Age distribution}
	Because of the difficulty in measuring accurate stellar ages, the distribution of stars in the [$\alpha$/Fe]-[Fe/H] plane is usually studied without consideration of stellar ages. 
	As discussed in the previous section, thanks to the combination of age determinations from spectroscopy and asteroseismology, stellar ages for large samples of stars beyond the solar neighbourhood have now become available. 
	In this paper we therefore aim at modelling age and chemical abundance {\it simultaneously}.
	
	In this section we discuss the observed distributions qualitatively and will then present the results from our chemical evolution model (described in Section~\ref{sec:model}) for a quantitative analysis in Section~\ref{sec:results}.
	
	\subsubsection{[Fe/H] vs [$\alpha$/Fe]}
	The left-hand panel of {Fig.}~\ref{afe-feh-age-thin} shows the distribution of the outer disc stars in the [Fe/H]-[$\alpha$/Fe] plane colour-coded by mean stellar age in each [Fe/H]-[$\alpha$/Fe] bin. 
	
	Two sequences with different ages can be identified separated by the dashed line: one sequence of young ($\sim 2\;$Gyr, blue colours) to intermediate-age ($\sim 4\;$Gyr, yellow colours) stars with systematically lower $[\alpha/{\rm Fe}]$ ratios by $\sim 0.05\;$dex and a sequence of intermediate-age to old ($> 7\;$Gyr, orange to red colours) stars with higher $[\alpha/{\rm Fe}]$. We note that a similar two-age-sequence in the [Fe/H]-[$\alpha$/Fe] plane for the low-$\alpha$ stars ($[\alpha/{\rm Fe}]<0.15$) is also seen in disc stars observed by the LAMOST survey (see Fig~4 in \citealt{wu2018}). Most surprisingly, the {\it younger} sequence (blue colours) is shifted towards {\it lower} [Fe/H], resulting an age-metallicity anti-correlation such that older stars are more metal-rich. 
	
	This trend is the opposite of what is expected from simple chemical evolution {with a single gas accretion phase}. 
	A single secular gas accretion phase is generally used to match the low-$\alpha$ sequence, such as the secular infall with long accretion timescales in the original 'two-infall' model by \citet{chiappini1997}. An important prediction of this secular evolution process is a monotonic metal enrichment history leading to element abundances increasing monotonically with time {(e.g., Figs.~13-17 in \citealt{anders2017})}. The presence of young metal-poor stars and old metal-rich stars is clearly in conflict with this model prediction. This trend suggests that the outer disc has experienced more than one major gas accretion event which perturbed its metal enrichment history. Most importantly, the young metal-poor stellar population must have formed in a metal-poor environment most plausibly caused by dilution with recently accreted, pristine gas.
	
	\subsubsection{Age vs [$\alpha$/Fe]}
	The right-hand panel of Fig.~\ref{afe-feh-age-thin} illustrates this further. Here we plot [$\alpha$/Fe] as a function of age colour-coded by [Fe/H]. There is no significant correlation between [$\alpha$/Fe] and age. But very interestingly, we notice the presence of a group of intermediate-age stars {(age$\sim5$\;Gyr)} with relatively high $\alpha$ abundance (0.1$<$[$\alpha$/Fe]$<$0.18) and low iron abundance ([Fe/H]$<$-0.3). 
	The presence of young $\alpha$-enhanced stars in a much smaller sample was already reported by \citet{martig2015} and \citet{chiappini2015}. A simple chemical evolution model cannot explain the formation of this stellar population.
	
	The relatively young age and high $\alpha$/Fe ratio of these stars suggest that they were formed in a recent, relatively short star formation episode. The low iron abundance further implies that this star burst must have been accompanied or triggered by the accretion of pristine/metal-poor gas. Note that the iron abundance of these stars is lower than the iron abundance of older {metal-rich ($[{\rm Fe}/{\rm H}]>0.1$)} stars by $\sim 0.5$ dex. 
	
	{Within this scenario, the most metal-rich stars ($[{\rm Fe}/{\rm H}]>0.1$, red box in the left-hand panel of Fig.~\ref{afe-feh-age-thin}) are the last population to form {\it before} the onset of dilution due to late accretion of metal-poor gas, while the most metal-poor stars ($[{\rm Fe}/H]<-0.4$, blue box in the left-hand panel of Fig.~\ref{afe-feh-age-thin}) are the last population to form in the accretion itself. The two boxes in the left-hand panel highlight these stars that possibly formed at the start and end point of the late-accretion event. The stellar populations in the metal-rich (red) and metal-poor (blue) boxes have average ages of $5.5\pm 3.1$ Gyr and $4.8\pm 1.9$ Gyr, respectively. This puts stringent constraints on the onset and duration of the late-accretion event.}
	
	In the following we use a numerical chemical evolution model to explore and quantify this scenario.
	
	
	\section{The chemical evolution model}
	\label{sec:model}
	The chemical evolution model used in this work was initially developed to understand the origin of the mass-metallicity relation for both gas and stars of local galaxies \citep{lian2018a}. The model was then expanded to explain the radial distribution of gas and stellar metallicity in local star-forming galaxies \citep{lian2018b,lian2019} as well as the cosmic evolution of the mass-metallicity relation from $z\sim 3.5$ to $z\sim 0$ \citep{lian2018c}. {The model is designed to be generic and can be used to infer the chemical evolution in galaxies globally as well as in sub-regions within galaxies. In this work we develop the model further to include the evolution of chemical element ratios to allow for a direct comparison with observations in our Galaxy.} We refer the reader to these papers for a detailed description of the model. Here we provide a brief summary of key aspects relevant to the current work.
	
	\subsection{Key parameters}
	The model considers three basic processes that regulate the chemical evolution of galaxies: star formation, gas accretion and galactic winds. The {initial gas accretion phase is assumed to decline exponentially.}  
	Two free parameters are used to describe the accretion phase, the initial accretion rate $A_{\rm acc,i}$ and the e-folding time-scale $\tau_{\rm acc}$. Star formation is implemented through the
	Kennicutt-Schmidt (KS) star formation law (SFL \citealt{kennicutt1998}), which implies the SFR surface density to be proportional to the gas mass surface density through a power law as described by the following equation.
	\begin{equation*}
	\Sigma_{\rm SFR}=2.5\times10^{-4}\times C_{\rm ks}\times(\Sigma_{\rm gas}/{\rm M_{\odot}pc}^{-2})^{n_{\rm ks}} {\rm M_{\odot}yr^{-1} kpc}^{-2}.
	\end{equation*} 
	We fix the power index $n_{\rm ks}$ to 1.5 for normal star forming galaxies \citep{kennicutt2007}, which is close to the original value of 1.4 obtained for star burst galaxies \citep{kennicutt1998}. The {parameter} $C_{\rm ks}$, which regulates the star formation efficiency {(SFE, star formation rate per unit gas mass)}, is normalised to the original {coefficient} of the KS law (i.e.\  $2.5\times10^{-4}$) and is allowed to vary. Different SFE values were found for different galaxies and also within galaxies (e.g., \citealt{leroy2008,zhang2019}). {The SFE may also increase with redshift. \citealt{lilly2013} find that an increase of the SFE with redshift explains the cosmic evolution of the mass-metallicity relation (however, see the recent review by \citealt{maiolino2019} for other possible mechanisms)}. 
	{A varying SFE is possibly related to the balance between the gas phases of HI and HII on small scales. The formation of HII from HI as well as its destruction depends on the local small-scale environment (e.g., stellar surface density and metallicity, \citealt{leroy2008}) which varies from galaxy to galaxy and even within individual galaxies.}
	
	Finally we note that
	{a Kroupa stellar initial mass function \citep{kroupa2001} is adopted in the model.}  
	
	\subsection{Nucleosynthesis prescription}
	Metal production from asymptotic giant branch (AGB) stars, Type-Ia supernovae (SN-Ia), and Type-II supernovae (SN-II) is included in the model, considering their different lifetimes. A detailed description of the yields table used in the model can be found in \citet{lian2018a}. The stellar yields of SN-II have been updated to the more recent work by \citet{kobayashi2006}. 
	
	To properly model [$\alpha$/Fe], the SN-Ia rate needs to be taken into account carefully.
	One of the methods in the literature adopts a `first-principle' approach to model the SN-Ia rate using a theoretical SN-Ia rate formalism \citep{greggio1983,matteucci1986,thomas1998}. This approach requires assumptions of the progenitor type of SN-Ia, the binary mass function, the secondary mass fraction distribution at a given mass of the binary system, and binary lifetimes. Because of the uncertainties in the nature of SN-Ia progenitors, empirical SN-Ia delay-time distributions (DTDs) calibrated to the observed SN-Ia rates have been proposed \citep{strolger2004,matteucci2006,maoz2012}. In this work we adopt the power-law SN-Ia DTD proposed by \citet{maoz2012} with slope $-1.1$ as follows: 
	\begin{equation*}
	{\rm DTD} = a(\tau/{\rm Gyr})^{-1.1}, 
	\end{equation*}  
	where $\tau$ is the delay time since the birth of the SN-Ia-producing binary systems, and $a$ is a normalisation constant {so that 
		\begin{equation*}
		\int_{\tau_{\rm min}}^{\tau_{\rm max}}{\rm DTD}(\tau)d\tau=1. 
		\end{equation*}
		Here $\tau_{\rm min}=\tau_{\rm 8\odot}$ and $\tau_{\rm max}=\tau_{\rm 0.85\odot}$, which are the minimum and maximum assumed lifetimes of a SN-Ia-producing binary, respectively. We will discuss the effect of different descriptions for the SN-Ia rate in \textsection5.1. }   
	
	\subsection{Models for the Milky Way disc}
	\begin{table*}
		\caption{Parameters adopted for {the model with a single accretion phase and a set of late-accretion models with different accretion timescales $t_{\rm d}$}. The subscript `i' stands for initial value of the parameter and `acc' stands for the parameter adopted during the late-accretion event. $C_{\rm ks,after}$ is the star formation law coefficient after the late-accretion event.} 
		\label{table1}
		\centering
		\begin{tabular}{l c c c c c c c c c }
			\hline\hline
			& $A_{\rm acc,i}$ & $\tau_{\rm acc,i}$ & $t_{\rm acc}$ & $t_{\rm d}$ & $C_{\rm ks,i}$ & $C_{\rm ks,acc}$ & $A_{\rm acc}$$^a$ & $C_{\rm ks,after}$ & ${\rm [Fe/H]}_{\rm acc}$ \\
			& ${\rm M_{\odot} yr^{-1} kpc^{-2}}$ & Gyr & Gyr & Gyr &  & & ${\rm M_{\odot} yr^{-1} kpc^{-2}}$  &  & dex \\
			\hline
			Single-$\tau$ & 0.002 & 10 & - & - & 0.50 & - & - & - & - \\	
			\hline
			$t_{\rm d}=0.05$Gyr  & 0.007 & 2 & 8.2 & 0.05 & 0.60 & 2.80 & 0.300 & 0.45 & pristine \\
			\hline
			$t_{\rm d}=0.7$Gyr (Fiducial)  & 0.007 & 2 & 8.2 & 0.7 & 0.60 & 0.85 & 0.024 & 0.14 & pristine \\ 
			\hline
			$t_{\rm d}=1.5$Gyr  & 0.007 & 2 & 8.2 & 1.5 & 0.60 & 0.60 & 0.020 & 0.10 & pristine \\
			\hline
			Continuous acc  & 0.007 & 2 & 8.2 & - & 0.60 & 0.60 & 0.024 & - & pristine \\
			\hline
			$t_{\rm d}=0.7$Gyr	& 0.007 & 2 & 8.2 & 0.7 & 0.60 & 0.85 & 0.036 & 0.14 & -1 \\ 
			\hline
		\end{tabular}\\  
	{Note $^a$: Constant accretion rate is assumed during the late-accretion event except for the continuous accretion model in which the accretion rate declines exponentially with an e-folding time of 8\;Gyr.} \\
	\end{table*}
	The success of our chemical evolution model at reproducing observations for a wide range in cosmic time and physical scale \citep{lian2018a, lian2018b, lian2018c} suggests that the key processes driving chemical evolution - star formation, gas inflow and outflow - have been adequately taken into account. In this paper we now expand this model further to constrain the detailed chemical enrichment history of our Galaxy through comparison with observations of individual stars. 
	
		\begin{figure*}
		\centering
		\includegraphics[width=14cm]{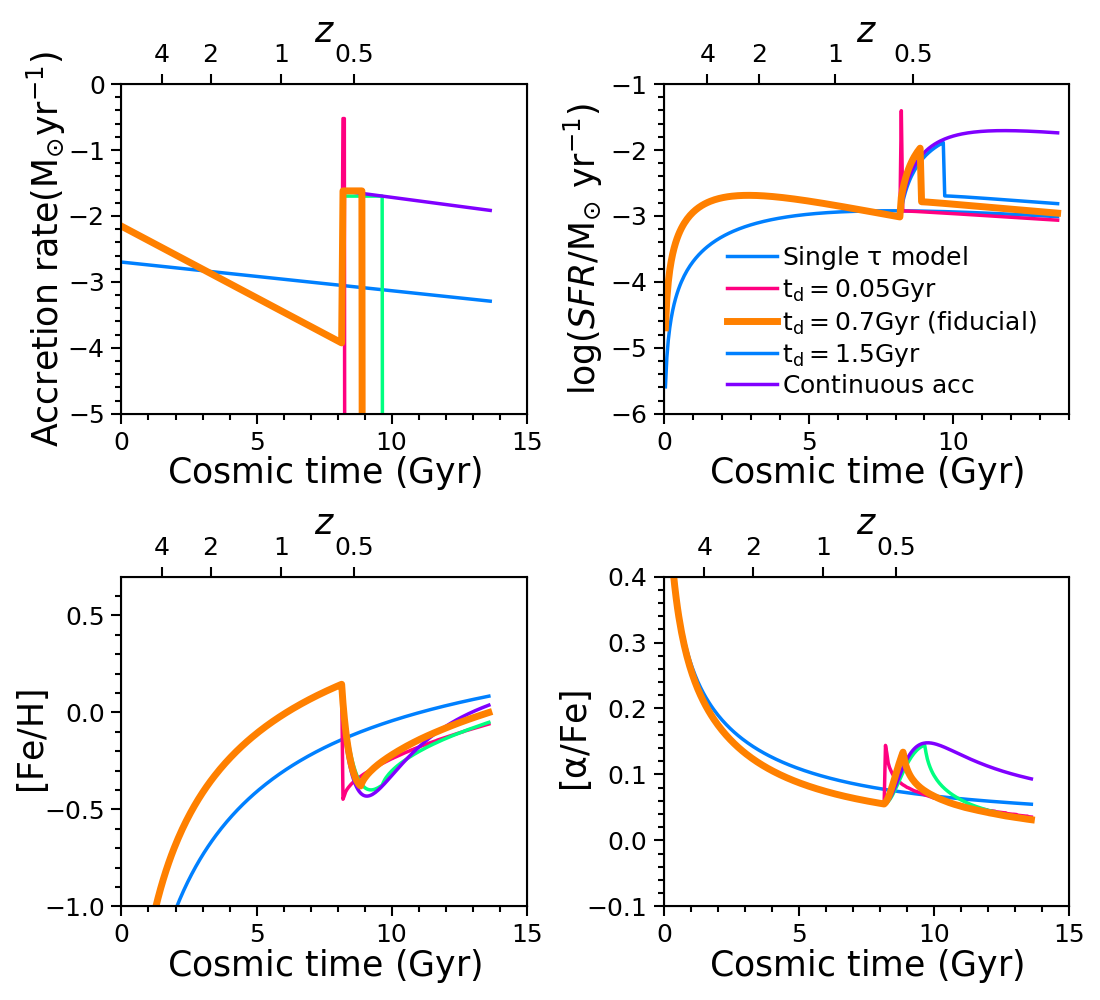}
		\caption{Gas accretion (top-left), star formation (top-right), iron abundance [Fe/H] (bottom-left), and [$\alpha$/Fe] abundance ratio (bottom-right) as a function of cosmic time for 
		{late-accretion models with different accretion timescales $\tau_{\rm d}$ as well as the single $\tau$ model (with single phase of gas accretion). Our fiducial model is a late-accretion model with a second infall timescale of 0.7\; Gyr.}	
		{The timescale of the secular gas accretion phase is 10\;Gyr in the {single-$\tau$ model} and 2\;Gyr in the late-accretion {models}. }
		The {late-accretion} {models are based on a modified single $\tau$ model} with an additional accretion event imposed on top of the secular accretion. The accretion decreases the [Fe/H] abundance, but the star burst which is triggered by the accretion still manages to enhance the [$\alpha$/Fe].}
		\label{sfh}
	\end{figure*} 

	By modelling gas and stellar metallicities of local star-forming galaxies we found that {the strength of metal outflow (i.e.\ the metal mass loading factor) depends on a galaxy's total stellar mass and plays a relatively unimportant role in the chemical evolution of} massive star forming galaxies {with stellar mass above $10^{10.5}\;{M_{\odot}}$} \citep{lian2018a,lian2018b}.  {We assume a today's stellar mass of $\sim 6\times 10^{10} M_{\odot}$ for our Galaxy \citep{mcmillan2011}. Therefore no metal outflow is assumed here for the Milky Way.} To keep the model generic, we calculate the chemical evolution within a squared area of size of 1 kpc$^2$. The size of the area can be considered as a normalisation factor of the model. 
	{All parameters in our model are {calculated per unit area and are therefore expressed in units of} kpc$^{-2}$. To get {results for a particular area, the values of our output parameters need to be scaled accordingly. The values for the entire outer disc are obtained by multiplication} with the area of the outer disc. For example, the initial gas accretion rate for the entire outer disc equals the initial accretion rate (0.007 M$_{\odot}{\rm yr^{-1}}{\rm kpc^{-2}}$ {in the late-accretion model}) times the area of the outer disc (392.7 kpc$^2$ for an annulus with radius between 10 and 15 kpc) which is 2.75 M$_{\odot}{\rm yr^{-1}}$. In the same way we estimate the total gas mass accreted in the late accretion event. We implicitly assume the outer disc to be homogeneous in stellar chemical abundances. This is supported by the spatially-resolved [$\alpha$/Fe]-[Fe/H] diagram in Fig.~4 in \citet{hayden2015} which shows insignificant radial variation within the outer disc. A more detailed study of abundance variation across the Galactic disc (e.g., the metallicity gradient) will be presented in future work.} 
	The time resolution of the model is 0.05 Gyr. 
	
	To perform a more direct comparison with the data, we simulate the stellar distribution predicted by the model after considering stellar mass loss.
	The observational uncertainties of [$\alpha$/Fe] and [Fe/H] are $\sim$0.02 dex on average \citep{ness2016}. We adopt these uncertainties and the age-dependent uncertainty of age from Table~\ref{tab:agescatter}. 
	
	
	As discussed in \textsection2.3 an additional phase of recent gas accretion on top of the secular accretion phase is required to explain the observed distribution of outer disc stars in the [$\alpha$/Fe]-[Fe/H]-Age space. 
	We explore this scenario quantitatively with a model adopting an additional short accretion event on top of the underlying secular accretion. 
	We refer to this model as `{late-accretion} model'. For comparison we also consider a model assuming a single exponentially declining accretion phase referred to as `{single-$\tau$} model'.
	
	In the {late-accretion} model, the additional accretion event is characterised by the {time of accretion $t_{\rm acc}$,} {the accretion timescale $t_{\rm d}$,} the {accretion rate} $A_{\rm acc}$, and an enhanced star formation efficiency achieved by an increased value for the coefficient of the KS law during the event $C_{\rm ks,acc}$.
	
	As discussed in \textsection2.3, the time of accretion $t_{\rm acc}$ is mostly constrained by the age of the observed {most metal-rich stars and the accretion timescale is given by the age difference between these and the most-poor stars. }
	The observational data imply the average age of {the most metal-rich} stars to be {5.5$\pm$3.1 Gyr and of the most metal-poor stars to be 4.8$\pm$1.9 Gyr}. This provides us with an estimate for $t_{\rm acc}$ {and $t_{\rm d}$ and their uncertainties}. 
	
	The accretion rate $A_{\rm acc}$ is constrained by the difference in [Fe/H] of the stars formed before and during accretion, i.e.\ the decrease in [Fe/H] due to dilution. The dilution effect is well constrained as the stars formed before and after the accretion event are observed.
	
	The enhanced coefficient of the KS law $C_{\rm ks,acc}$ is largely constrained by the enhancement of [$\alpha$/Fe] during the accretion event ($0.06<[\alpha/{\rm Fe}]<0.13\;$dex). {As we will show later, less enhancement in SFE is required in case of a longer accretion timescale.} Table~\ref{table1} summarises the model parameters.
	
	The parameters adopted for the {single-$\tau$ model and the late-accretion models with various $t_{\rm d}$} {for the outer disc} are listed in Table~\ref{table1}. The additional late accretion in the {fiducial  late-accretion model} occurs at a look-back time of {5.5} Gyr corresponding to a cosmic time of {8.2} Gyr or $z\sim 0.6$. {$1.68\times10^7{\rm M_{\odot}/kpc^{2}}$} pristine gas is accreted. {For comparison, \citet{spitoni2019} adopt an earlier second gas accretion at a larger look-back time of 9.4\;Gyr (or cosmic time of 4.3\;Gyr).}
	
	The coefficient of the star-formation law is enhanced by a factor {1.4}. Based on the age difference between the most metal-rich and the most metal-poor populations, we adopt an accretion timescale of 0.7 Gyr in the {fiducial} late-accretion scenario, assuming a constant accretion rate during accretion.
	
	{We test this choice by exploring various late-accretion models with different accretion timescales (see Section 4 and Figs.~\ref{sfh}, \ref{track}). It should also be noted that the observed peak in [$\alpha$/Fe] can be matched with lower SFL coefficient in models with longer accretion timescales. For an accretion with a timescale as long as 1.5 Gyr, an enhancement in SFL coefficient is no longer needed. This is because a larger total amount of gas has to be accreted in models with longer accretion time-scales, in order to reach the same level of net dilution because of the additional chemical enrichment during the accretion episode itself.}
	
	{The final stellar mass surface density and gas mass fraction of the {fiducial} late-accretion model are 12.3 M$_{\odot}/{\rm kpc}^2$ and 41\%, respectively. 
	{Assuming a disc scale-length of 2.6 kpc and a solar stellar mass surface density of 33.4 ${\rm M}_{\odot}$pc$^{-2}$, the stellar mass surface densities at the galactocentric radii of 10 and 15$\;$kpc are 15.5 and 2.3 ${\rm M}_{\odot}$pc$^{-2}$, respectively. Adopting the radial distribution of the gas mass surface density from \citet{wolfire2003}, we estimate the gas mass fraction (i.e.\ $\frac{M_{\rm gas}}{M_{\rm gas}+M_{\rm star}}$) within the radii of 10 to 15$\;$kpc to be $28-68$ per cent. Our results are well in line with these estimates.}	}
	
	
	\section{Results}
	\label{sec:results}
	In this section we describe in detail the predictions of our models and compare them to observations.
	
	\subsection{Predicted evolutionary histories}  
	
		\begin{figure*}
		\centering
		\includegraphics[width=15cm]{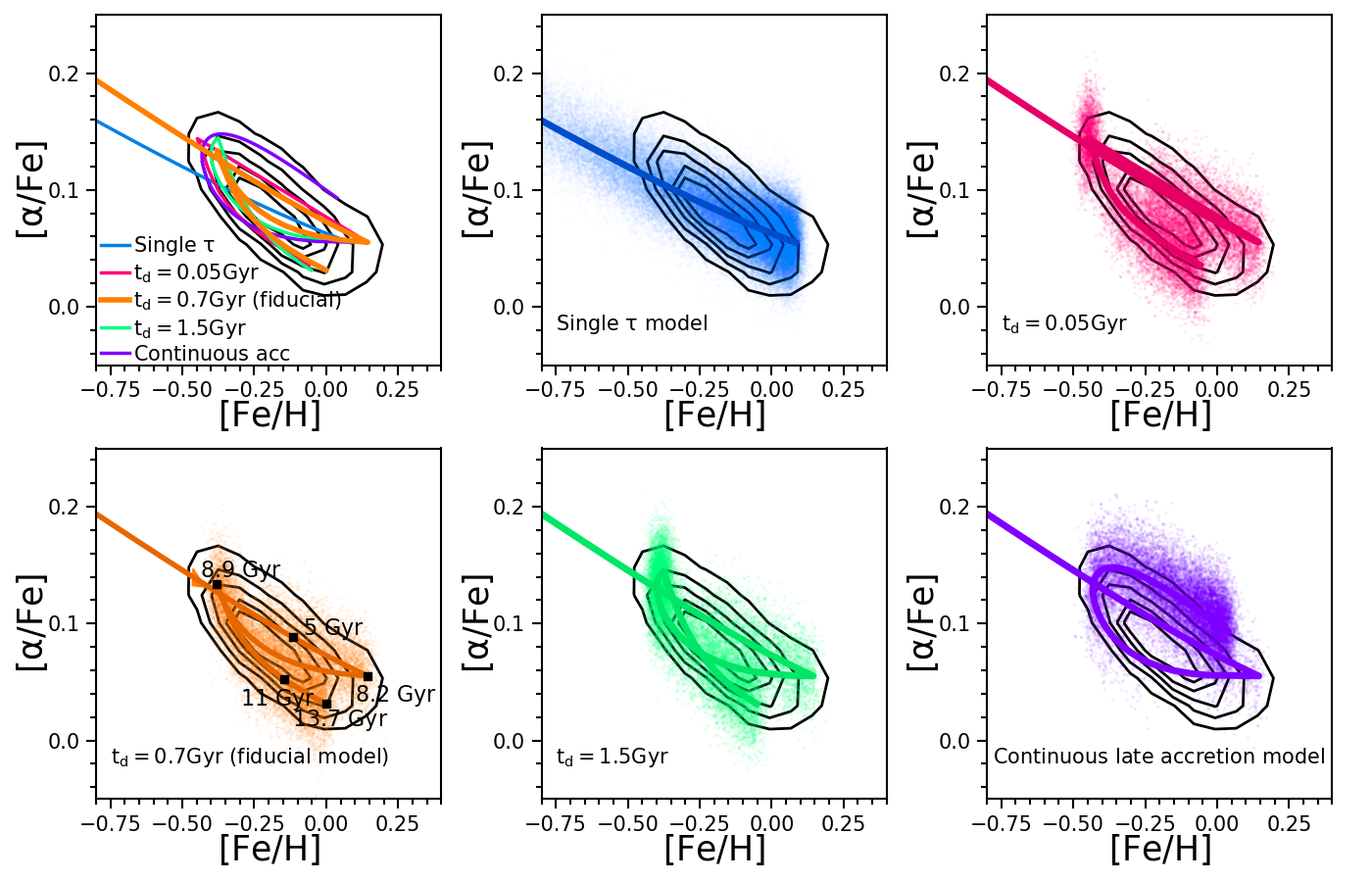}
		\caption{Comparison of model predictions with observational data in the [$\alpha$/Fe]-[Fe/H] plane. The black contours show the observed distribution while the lines are the model tracks with the same colour scheme as in Fig~\ref{sfh}. 
		The coloured dots in each panel are the simulated distributions after considering observational uncertainties. 
		{Five epochs at 5, 8.2, 8.9, 11, and 13.7\;Gyr} in the {fiducial} {late-accretion} model are marked as blacks squares {for reference}.}
		\label{track}
	\end{figure*}
		Figure~\ref{sfh} shows the evolution of gas accretion rate (top-left panel), star formation rate (top-right panel), [Fe/H] (bottom-left panel), and [$\alpha$/Fe] (bottom-right panel) as predicted by {various late-accretion models with different accretion timescales $t_{\rm d}$ as well as the single accretion model. We adopt an exponentially declining timescale of 8 Gyr for the continuous late-accretion which is taken from \citet{spitoni2019}.}
	
	The {late-accretion} {models are} based on {a secular accretion phase} with $\tau_{\rm acc,i}=2\;$Gyr 
	to which a second, {late} gas accretion event is added on top. {The initial secular accretion phase is not well constrained due to lack of observed old stars with age$\;>8$\;Gyr which formed during this phase in the outer disc. As a consequence, there is a degeneracy between the parameters characterising this phase, including the initial gas accretion rate, the accretion timescale, and the initial SFE. {This degeneracy can be broken using observations in the inner Galaxy for which observational data of old stars are available.} An improved model using data from the inner discs to constrain this early accretion phase will be presented in companion papers on the inner disc and bulge (Lian et al. in prep.).} 
	
	In our model, the {initial secular and the late} accretion events are independent of each other. The late-accretion event, unlike the underlying smooth accretion, is designed to occur abruptly on a short timescale {(0.7 Gyr in our fiducial model)}. This is required to boost the [$\alpha$/Fe] in the ISM.
	With the combination of an enhanced star formation efficiency, a star burst is induced for this accretion event (top-right panel of Figure~\ref{sfh}).  
	
	This late-accretion event affects the chemical enrichment significantly. Individual element abundances, such as [Fe/H], are diluted dramatically, resulting in a sharp drop in its abundance (bottom-left panel of Figure~\ref{sfh}).
	For the same reason the $\alpha$-element abundances also drop initially by the same factor. However, because of the star burst which is triggered at the same time, the $\alpha$-abundance is boosted through chemical enrichment from short-lived Type-II supernova. As a result, the [$\alpha$/Fe] ratio peaks when the accretion event occurs (bottom-right panel of Figure~\ref{sfh}).
	
	To summarise, there are three majors effects induced by the late-accretion event: 1) a burst of star formation, 2) a suppression of iron abundance, and 3) an enhancement of the [$\alpha$/Fe] abundance ratio. These features create the key difference between the predictions of the {late-accretion} {models} and the {single-$\tau$} model. 

	\subsection{[$\alpha$/Fe]-[Fe/H] diagram}
	
	\begin{figure*}
		\centering
		\includegraphics[width=18cm]{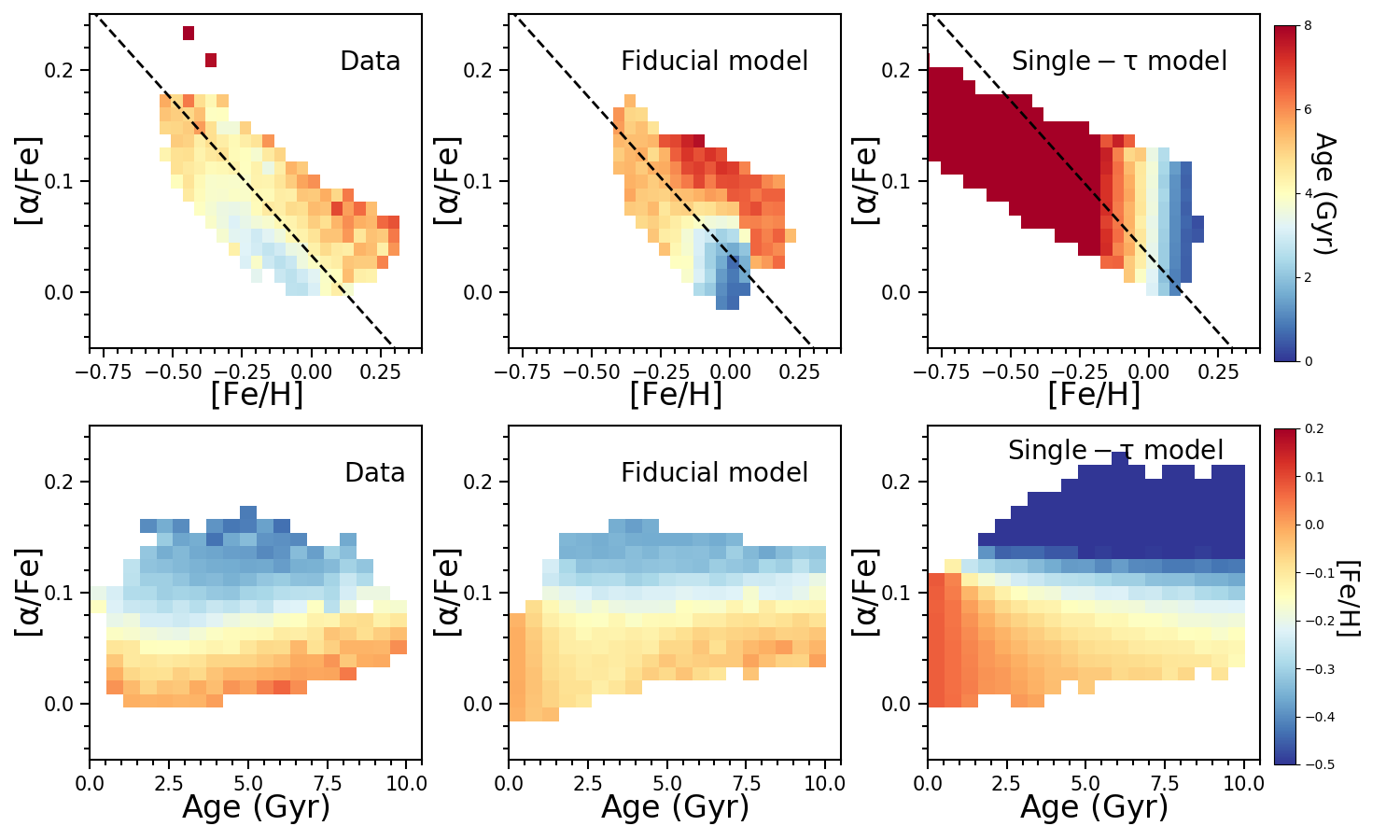}
		\caption{Comparison of model predictions with observational data in the [$\alpha$/Fe]-[Fe/H] plane colour-coded by mean age (top row) and in the [$\alpha$/Fe]-Age plane colour-coded by mean [Fe/H] (bottom row). Columns from left to right are: observations, {late-accretion} model and single $\tau$ model, respectively. The black dashed line in the top panels separates the two sequences in the [$\alpha$/Fe]-[Fe/H] plane (see Fig~\ref{afe-feh-age-thin}).}
		\label{afe-feh-age-model}
	\end{figure*}
	
	
	Figure~\ref{track} shows the comparison of the {single-$\tau$} and {late-accretion} models with observations in the [$\alpha$/Fe]-[Fe/H] plane. The black contours indicate the observed data while the coloured lines are the predicted model tracks. The colour scheme is the same as in Fig~\ref{sfh}. 
	{The top-left panel summarises the evolutionary trajectories of the different models while other panels visualise the predicted distributions of the stellar abundances for each model (coloured dots) after considering the observational uncertainties. An uncertainty of 0.02 dex for [$\alpha$/Fe] and [Fe/H] is adopted. The age uncertainty is summarised in Table\ref{tab:agescatter}.} 
	The beginning and the end of the late gas accretion episode in the {fiducial} {late-accretion} model are marked with black squares. 
	
	It can be seen that the {single-$\tau$} model {track} 
	appears to match the observed {trend} equally well as the {late-accretion} {models} in the [$\alpha$/Fe]-[Fe/H] plane. This model is similar to the models generally used in the literature to explain the chemical compositions of thin disc stars \citep[e.g.,][]{chiappini2009,anders2017}. 
	{This is challenged, however, by the distribution of stellar abundances in this diagram which peaks at a lower [Fe/H] and appears to be more symmetric. It can be seen that the latter is best matched by the fiducial late-accretion model (bottom-left panel).}
	
	{It is also interesting to note that the distribution in [$\alpha$/Fe]-[Fe/H] as predicted 
by the late-accretion model is actually composed of two parallel evolutionary sequences, one before and one after the accretion event}. The first sequence consists of stars formed from gradually accreted gas prior to the accretion episode and the other is made of stars formed during and after the late accretion. These two [$\alpha$/Fe]-[Fe/H] sequences of stars with different ages are in good agreement with the observed age-chemical structure shown in Fig~\ref{afe-feh-age-thin}. 
	
	\subsection{[$\alpha$/Fe]-[Fe/H]-Age relation}
	In the previous section we compared model and data in the [$\alpha$/Fe]-[Fe/H] diagram without age information. {Here we add age as a further parameter as shown in Figure~\ref{afe-feh-age-model}. 
	For simplicity, we only include the single-$\tau$ and the fiducial late-accretion model.} The top row shows the [$\alpha$/Fe]-[Fe/H], and the bottom row the [$\alpha$/Fe]-Age planes, respectively (see Fig~\ref{afe-feh-age-thin}).
	The left-hand column shows the observational data, the middle and right-hand columns show the predicted distributions from the {fiducial} {late-accretion} model and the {single-$\tau$} model, respectively.
	The black dashed lines in the top panels separate the two [$\alpha$/Fe]-[Fe/H] sequences with different ages (see Fig~\ref{afe-feh-age-thin}).
	
	
	
	It can be seen that the observed distribution in [$\alpha$/Fe]-[Fe/H]-Age is well reproduced by the {fiducial} {late-accretion} model.
	The two [$\alpha$/Fe]-[Fe/H] sequences with different ages seen in the observations (top-left panel) are well matched (top-middle panel).
	In particular, the observed age-metallicity anti-correlation visible through the presence of old metal-rich and young metal-poor stars is matched by the {fiducial late-accretion} model. 
	
	This model also reproduces the presence of intermediate-age ($2<t<6\;$Gyr), metal-poor ($[{\rm Fe/H]}<-0.25$) stars with relatively high [$\alpha$/Fe] ratios {([$\alpha$/Fe$]\sim0.14$)} as shown in the bottom panels. 
	As discussed above, these stars are the first generation formed during the star burst triggered by the late-accretion event.
	
	The {single-$\tau$} model (right-hand panels), instead, fails to reproduce the complex age-element abundance structure.
	As ought to be expected, this model predicts [Fe/H] to increase monotonically with decreasing age (hence increasing cosmic time), accompanied by a decreasing [$\alpha$/Fe] ratio. As a result, only a single evolutionary sequence is present in the [$\alpha$/Fe]-[Fe/H] plane. All stars with high [$\alpha$/Fe] are predicted to form at early times with old ages, in contrast to what is observed. 
	
	To summarise, the two-phase disc formation scenario with a delayed second accretion event as described by the {fiducial late-accretion} model provides a much better fit to the observed age-element abundance structure of the outer disc. The simple model based on a single phase of secular evolution fails.



	\section{Discussion}
	{In this section we will discuss uncertainties, implications, and predictions of our late-accretion model and confront with alternative scenarios in the literature.} 
	
	\subsection{The late-accretion model}
	
	\subsubsection{The effect of different SNIa delay time distributions}
	
	\begin{figure}
		\centering
		\includegraphics[width=8.5cm]{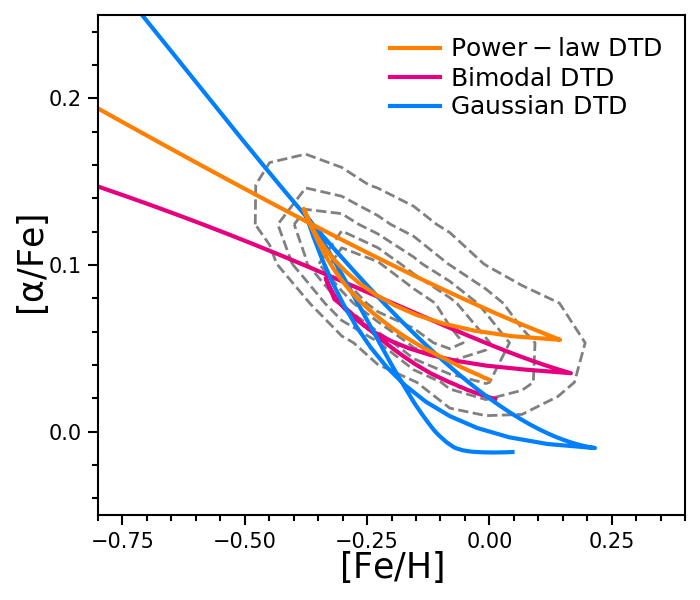}
		\caption{Late-accretion models with different descriptions for the SN-Ia rate based on different delayed-time distributions (DTD). Parameters in these models are identical as listed in Table~\ref{table1}. }
		\label{afe-sn1a}
	\end{figure}
	
	{It is easy to understand that the evolution of [$\alpha$/Fe] is sensitive to the description of the SN-Ia rate which dominates the production of iron. Figure~\ref{afe-sn1a} shows models with the same parameter configuration as before but with different DTDs (power-law DTD, \citealt{maoz2012}; bimodal DTD, \citealt{matteucci2006}; Gaussian DTD, \citealt{strolger2004}). It can be seen that the [$\alpha$/Fe] values are affected by up to $\sim 0.1\;$dex at a given [Fe/H]. Nevertheless, the overall evolutionary trends in the [$\alpha$/Fe]-[Fe/H] plane as predicted by these different models are broadly consistent. Although the parameters adopted in our late-accretion model are subject to small changes for different DTDs, the main result of this paper, namely that a late accretion event is required to explain the age-chemical abundance structure of the outer disc, remains unchanged. Moreover, it worth pointing out that some key parameters that characterise the late-accretion event are determined directly by the observations and are therefore not affected by changes of the DTD or other model configurations, including the onset time (8.2\;Gyr) and timescale (0.7 Gyr) of the late-accretion episode. Likewise, the fraction of pristine gas accreted is well constrained by observations.} 

	\subsubsection{Late accretion: a galaxy merger?}
	In the present paper we show that an additional, {short} accretion episode {delayed by $\sim$8.2\;Gyr} is required to explain the complex age-element abundance structure of the Galactic outer disc. 
	Our model suggests that the accreted gas is not connected to the underlying secular gas accretion. 
	We further find that the mass of accreted gas needs to exceed the mass of the local gas reservoir in the outer disc by a factor three. These characteristics {disfavour quasi-continuous gas accretion from the circumgalactic medium and suggest that the late-time gas accretion may be caused by the accretion of a gas-rich dwarf galaxy. As in a minor merger, the accretion of gas generally happens earlier than the accretion of stars through ram pressure processes, the stellar component of the infalling galaxy may not be fully disrupted yet. 
	Interestingly, recent simulations by \citet{tepper2018} suggest that the massive stripping of gas from the infalling Sagittarius dwarf galaxy (\citealt{ibata1994}) started $\sim 3$ Gyr ago, which makes Sagittarius a promising candidate for being the accreted galaxy leading to the late-accretion episode discussed in this paper.}
	 
	The model presented here allows us to infer several key properties of the accreted galaxy. Since the lowest iron abundance of the stars formed during the accretion event is $\sim -0.5\;$dex, the iron abundance in the gas of the accreted galaxy {would have to be} lower than $-0.5$ dex to be able to dilute the gas in the outer Galactic disc. 
	This upper limit in metallicity can be translated to an upper limit in {stellar} mass {of the accreted galaxy}. Using the observed {(stellar)} mass-{(gas-phase)} metallicity relation of galaxies at intermediate redshift ({i.e. $z\sim0.3-0.7$}, \citealt{savaglio2005,lian2016}) we find this upper mass limit to be $M_*\sim 10^9{M_{\odot}}$. {The stellar mass of Sagittarius is estimated to be a few times $10^8{M_{\odot}}$ \citep{gibbons2017}, which is well consistent with the upper limit estimated here.} 

	Assuming that the gas (and the subsequently formed stars) of the accreted galaxy settled evenly in the outer disc around $10 < r < 15$ kpc (the region we are analysing in the present paper), the required {\em gas} mass of the accreted galaxy {would have to be} $3\times 10^{9}{M_{\odot}}$. Combining this with the estimate of the {\em stellar} mass based on the mass-metallicity relation above, the gas fraction of the accreted galaxy {is estimated to be} $\sim 75$ per cent. {The merger hypothesis advocated here would therefore imply the system accreted by our Galaxy to be a gas-rich dwarf galaxy}. 	
	
		\begin{figure*}
		\centering
		\includegraphics[width=16cm]{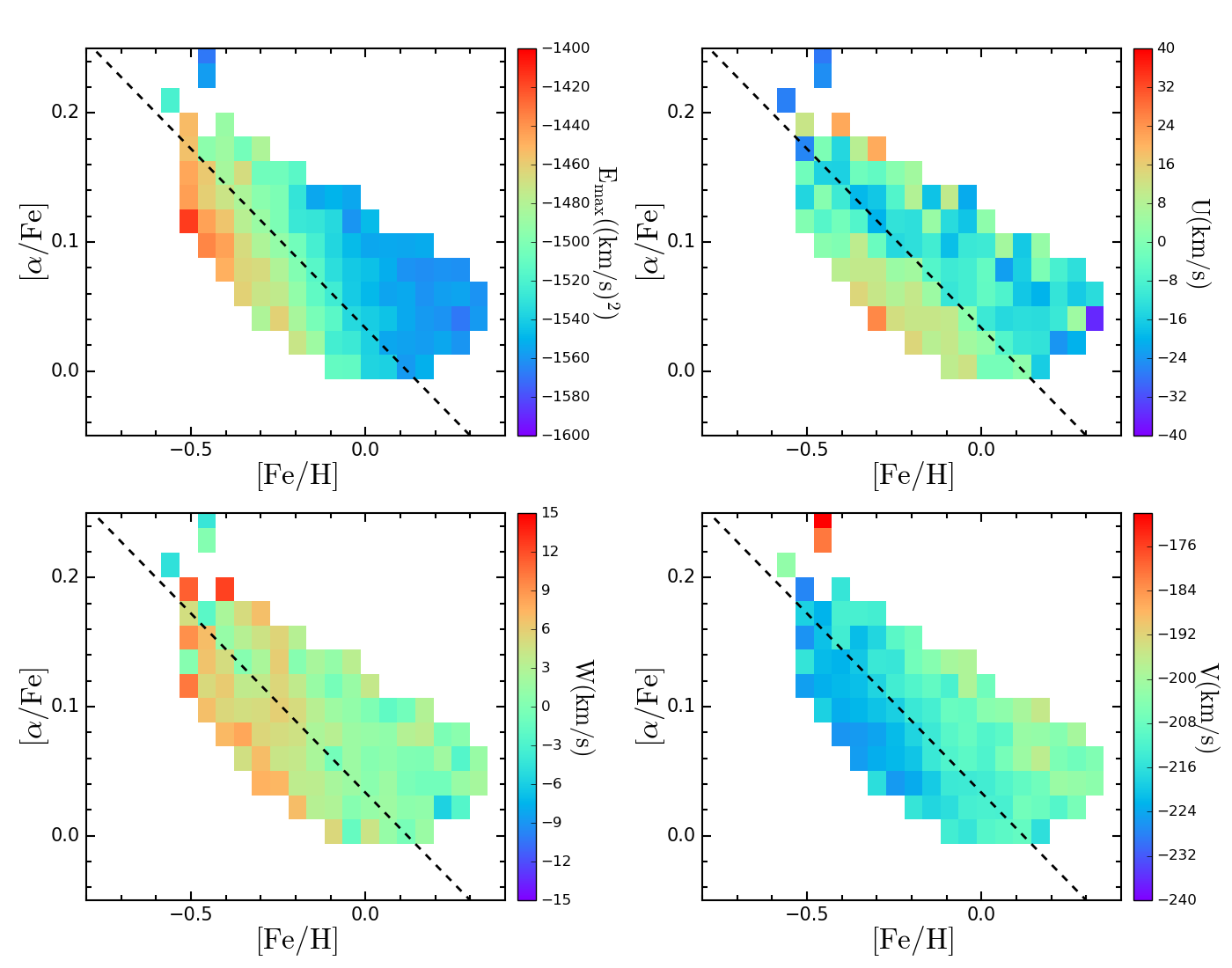}
		\caption{Distribution of the kinematic properties of stars in the [$\alpha$/Fe]-[Fe/H] plane. {\it Top-left panel}: maximum total orbit energy. {\it Top-right panel}: radial velocity $U$. {\it Bottom-left panel}: vertical velocity $W$. {\it Bottom-right panel}: rotation velocity $V$.}
		\label{kinematic}
	\end{figure*}	

	\subsubsection{Disc flaring}
	{The thickness of the Galactic disc {(i.e.\ its scale length)} is generally found to be {constant with radius} \citep{bovy2012a,bovy2012b}. However, this is not the case for sub-populations in the disc. The scale height of the young sub-component increases with increasing radius, which is usually referred to as `flaring' \citep{kalberla2014,carraro2015}. 
		By analysing the spatial structure of mono-abundance populations, \citet{bovy2016} and \citet{mackereth2017} systematically studied thickness and flaring as a function of [$\alpha$/Fe] and [Fe/H]. It turns out flaring is complex and depends chemical abundances. In general, the flaring amplitude is negligible for the high-$\alpha$ population but significant for the low-$\alpha$ population. Moreover, within the low-$\alpha$ population, flaring is most pronounced for the most metal-poor population with ${\rm [Fe/H]}\sim -0.5$ dex. 
		
		Radial migration is one of the mechanism believed to be responsible for flaring. 
		The idea is that the stars moving outward face less gravitational pull in the outer disc and therefore move to larger vertical distances. However, to explain the dependence of flaring on the element abundances as discussed above, a contrived mechanism for radial migration would be required. 
		
		The {late-accretion} model presented here provides an alternative scenario for explaining the flaring in the low-$\alpha$ population. Our model implies that the most metal-poor, low-$\alpha$ population forms during a recent, merger-induced star burst. The large scale-height of this population could then well be caused by the hot kinematics of the gas that fuels the star burst.  A larger vertical scale height of this population in the outer disc is indeed expected, if the strength of the star burst and the associated gas accretion is more pronounced at large radii. This would lead to the observed flaring pattern.}  
		
	\subsubsection{Kinematics}
	{In the following we briefly investigate the observed kinematic properties of stars in the outer disc in light of the proposed {late-accretion} model. Figure~\ref{kinematic} shows the distribution of four kinematic properties (maximum total orbit energy $E_{\rm max}$, radial velocity $U$, vertical velocity $W$, and rotation velocity $V$) in the [$\alpha$/Fe]-[Fe/H] plane. These kinematic parameters are taken from a synergistic catalogue of APOGEE and {\sl Gaia}.
	
	The orbit information of each star is derived with the GravPot16  code\footnote{https://gravpot.utinam.cnrs.fr} based on a {3-D steady-state gravitational potential model for the Galaxy, modelled as the sum of axisymmetric and non-axisymmetric components (Fernandez-Trincado et al., in prep). The axisymmetric component is made up of the superposition of many composite stellar populations from the thin disc and a contribution from the thick disc. The density profile of the thin disc component is assumed to follow the Einasto laws \citep{einasto1979}, while a sech$^{2}$ law is adopted for the density profile of the thick disc component. The non-axisymmetric component is modelled by a “boxy/peanut” bar structure whose density profile is observationally constrained from 2MASS data (see \citealt{robin2012}). These stellar components are assumed to be surrounded by an isothermal dark matter halo component with a mass density as presented in \citet{robin2003}.
	
	For the computation of Galactic orbits, we employ a simple Monte Carlo approach and the Runge-Kutta algorithm of seventh-eight order. For each APOGEE star, a thousand orbits are computed backward in time during 3 Gyr. As to the input parameters, we use the sky positions and line-of-sight velocities from the APOGEE survey. The proper motions and spectrophotometric distances are adopted from Gaia DR2. The spectrophotometric distances from \texttt{StarHorse} are relatively precise even out to large distances.} 
		
	Within the low-$\alpha$ sequence, the metal-poor populations (below the dashed line) and the metal-rich populations (above the dashed line) show systematically different kinematic properties. The total orbit energy in the metal-poor stars is generally higher than the total orbit energy in the metal-rich stars, which is mainly caused by their faster rotation velocities. The metal-poor stars also tend to have non-zero radial and higher vertical velocities than the metal-rich stars. All these differences suggest that the metal-poor populations likely formed from gas with hotter kinematics. This is qualitatively in line with the dynamical effect of a {gas-rich merger event. The latter is expected to lead to vigorous star formation in a dynamically hot environment with complicated kinematics \citep{mihos1998,colina2005}}.
	
	Interestingly, a merger event between the Milky Way and a dwarf galaxy named Gaia-Enceladus was recently identified \citep{helmi2018}. This discovery is based on the peculiar velocity distribution of stars accreted from the dwarf galaxy with respect to the normal disc stars in the Milky Way.}
	
	\begin{figure}
		\centering
		\includegraphics[width=8.5cm]{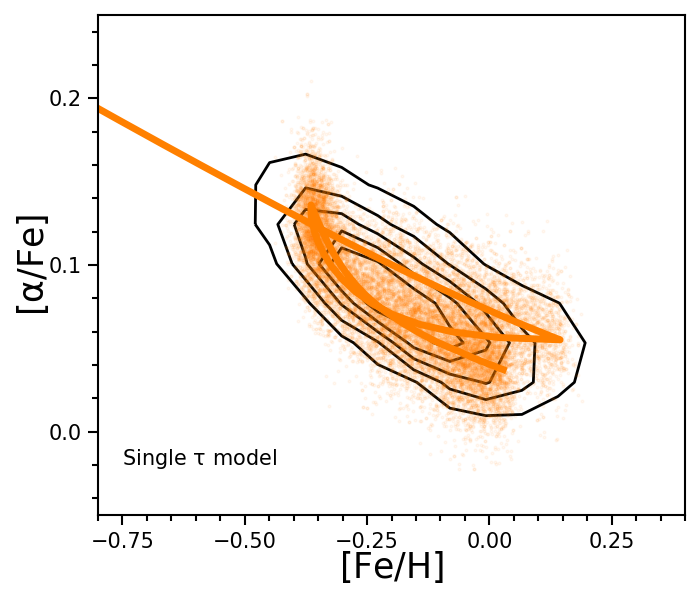}
		\caption{The same as {bottom-left panel of} Figure~\ref{track} but assuming the additional gas accretion has been enriched to 10\% of solar abundance.}
		\label{track-enrich}
	\end{figure}
	
	\subsubsection{Non-pristine accretion}
	In our {ficudial late-accretion} model we assume the additional accreted gas to be pristine. We now
	explore whether and how our results will be affected if the accretion is already metal-enriched to some extent. 
	To this end, we assume ${\rm [Fe/H]}=0.1\;$dex for the accreted gas with a solar element abundance pattern. The comparison between the model prediction and observations in the [$\alpha$/Fe]-[Fe/H] plane is shown in Figure~\ref{track-enrich}. It can be seen that the model prediction matches the data equally well. The best-fit parameters adopted for this model are listed in Table~\ref{table1}. Compared to the {fiducial} model with pristine gas accretion, more gas accretion ({50} per cent) is needed in order to dilute the iron abundance to the same value as before.

	\subsection{Alternative scenarios}
	In the following we discuss other scenarios proposed in the literature.
	
	\subsubsection{Radial migration}
	\begin{figure*}
		\centering
		\includegraphics[width=18cm]{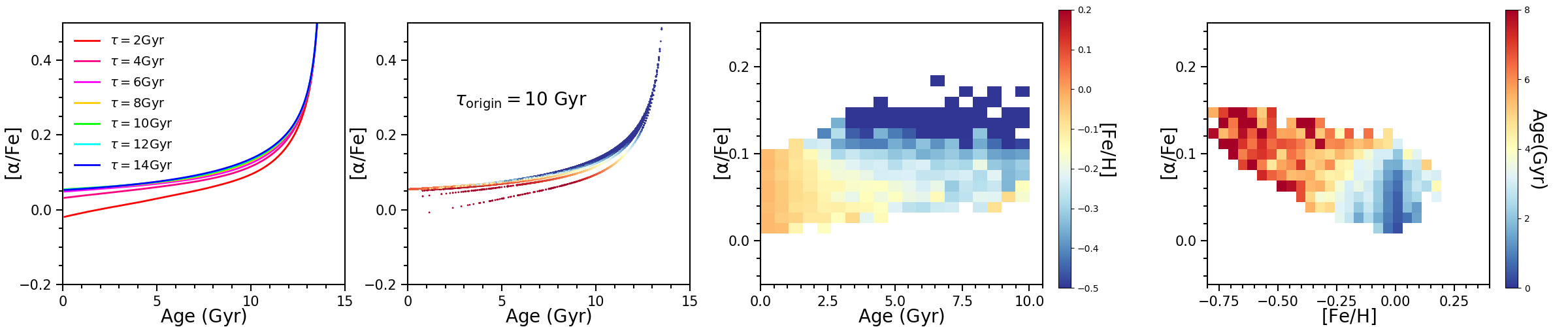}
		\caption{Theoretical predictions of the single-$\tau$ model including effects of radial migration. {\it Left-hand panel}: [$\alpha$/Fe] as a function of age showing seven models with different gas accretion timescales from 2 to 14 Gyr for different radial bins. The shortest timescale is used for the inner-most radial bin and increases with radius. {\it Centre-left panel}: Predicted distribution in the [$\alpha$/Fe]-Age plane colour-coded by [Fe/H] for the model of the radial bin at 11\;kpc including the effects of radial migration (see text for details). {\it Centre-right and right-hand panels}: Predicted distribution in [$\alpha$/Fe]-Age and [$\alpha$/Fe]-[Fe/H], respectively, after considering observational uncertainties and selection effects. }
		\label{migration}
	\end{figure*} 
	
	The main reason why the single-$\tau$ model fails in matching observations is that  element abundances increase with time in this model. This leads to a monotonic age-metallicity relation.
	This prediction is in disagreement with the observed complex age-element abundance structure of the Galactic outer disc. Most importantly, the coexistence of young, metal-poor and old, metal-rich populations cannot be understood with monotonic chemical enrichment.
	
	The APOGEE data show that the stars in the Galactic inner disc are generally older and more metal-rich than the stars in the outer disc. Still, the outer disc also contains a fraction of old, metal-rich stars very similar to the ones on the inner disc \citep{hayden2015}. Hence, it may be possible that these stars originally formed in the inner disc and migrated to the outer disc over time \citep{feuillet2018}. Such radial migration is indeed seen in simulations and is thought to play an important role in re-distributing stars in the Milky Way \citep{schonrich2009,minchev2013}. 
	
	To test whether radial migration can explain the observed age-element abundance pattern in the outer disc, we run a version of the single-$\tau$ model in which the effect of radial migration is included. To model the chemical enrichment at different radial bins, we follow the approach of \citet{chiappini2001}. We assume a single gas accretion phase with the accretion timescale increasing linearly with increasing radius. We calculate a set of seven models with the gas accretion timescale ranging from 2 to 14 Gyr. According to \citet{chiappini2001}, this corresponds to a range in radius from 3 kpc to 15 kpc. {The decrease in stellar mass density with increasing radius is considered in these models.} The evolutionary tracks of age vs [$\alpha$/Fe] as predicted by the resulting model is shown in the left-hand panel of Figure~\ref{migration}.
	
	We then include the effect of radial migration by implementing a mass exchange between adjacent radial bins. At each time step, a certain fraction of stars is set to move toward adjacent radial bins both out- and inward. The time step assumed here is 1 Gyr and the fraction is 20 per cent. We verified that the results do not depend much on the exact values of these two parameters. The radial migration fraction of 20 per cent per Gyr is relatively high and broadly consistent with cosmological simulations \citep{ma2017}.
	
	The centre-left panel of Figure~\ref{migration} shows the resulting distribution in the [$\alpha$/Fe]-Age plane colour-coded by [Fe/H] for the radial bin {with an accretion timescale of $\tau=10$\;Gyr. This roughly corresponds to a radius of $11 {\rm kpc}$}. 
	The centre-right and right-hand panels show the final distributions in the [$\alpha$/Fe]-Age and [$\alpha$/Fe]-[Fe/H] planes, respectively, taking into account the observational uncertainties.
	
		
	This model does not perform significantly better than the original single-$\tau$ model. The monotonic increase of [$\alpha$/Fe] and [Fe/H] with cosmic time is still clearly visible.
	Even though radial migration adds some fraction of old metal-rich stars from the inner disc to the outer disc, which perturbs {the distribution in [$\alpha$/Fe]-Age and [$\alpha$/Fe]-[Fe/H], the improvement is marginal.
		
		
		We conclude that radial migration mitigates but does not fully resolve the difficulties of the single-$\tau$ model in matching the observational data.} 
	
	{Further evidence against radial migration playing a major role comes from the radial distribution. Assuming {a constant radial migration efficiency with time, a larger scale length with a wider radial distribution would be expected for older populations simply because they have had more time to migrate}. However, \citet{bovy2012b,bovy2016} {find} the scale-length for the {old} metal-rich stars in the low-$\alpha$ sequence to be small ($\sim2\;$kpc). This scale {length} is similar to the one of the {old} stars in the high-$\alpha$ sequence but systematically smaller than the one of the {younger} metal-poor populations in the low-$\alpha$ sequence ($\sim4\;$kpc).}
	
	\subsubsection{The modified two-infall model}
	\citet{spitoni2019} developed a modification of the classical `two-infall' model in parallel to the present work. Similar to our late-accretion model, their model proposes a second phase of gas accretion delayed by 4.3 Gyr. This second accretion phase is responsible for the formation of low-[$\alpha$/Fe] stars. This modified `two-infall' model is indeed able to reproduce the observations better. Indeed this model is similar to our {late-accretion} model in that the second phase of gas accretion is delayed by several Gyr. However, fundamental {differences} remain.
	
	{First, the delayed accretion in our model is pushed to much later epochs than in \citet{spitoni2019} (8.2\;Gyr versus 4.3\;Gyr). The earlier accretion in the \citet{spitoni2019} model results in a {slight} overestimation of the age of the low-[$\alpha$/Fe] stars.}
	{Second, the} model by \citet{spitoni2019} considers continuous gas accretion during the second delayed accretion phase with {an} exponentially declining timescale of $8\;$Gyr. The {late-accretion} model proposed here, instead, considers a {short burst of star formation with a timescale of $\sim 0.7\;$Gyr} for this second accretion phase. This is required to match the {age difference between the stars formed prior to and at the end of the late-accretion episode. } 
	
	\subsubsection{A heuristic scenario}
	Also in parallel to the present work, \citet{haywood2019} developed a heuristic scenario to explain the age-element abundance structure of the Galactic disc. 
	In this scenario, the disc is pre-enriched by the formation of a high-[$\alpha$/Fe] population (i.e.\ the chemical thick disc) and then diluted by metal-poor gas inflow from the outer disc which fuels a second phase of star formation and forms the chemical thin disc. 
	This model provides an interesting picture of the thick/thin disc formation but requires further refinements to match the detailed distribution of stars in the [$\alpha$/Fe]-[Fe/H]-Age space. 
	For example, one of the challenges faced by this scenario is the observed coexistence of young, metal-poor and old, metal-rich populations {in the low-$\alpha$ sequence}. 
	
	\subsubsection{Delayed accretion in cosmological models}
	{It is interesting to note that delayed gas accretion has already been considered in hydrodynamical simulations \citep{birnboim2007} and semi-analytic models \citep{calura2009}. In fact the presence of a flat low-$\alpha$ sequence that resembles the now observed low-$\alpha$ sequence but with much lower [$\alpha$/Fe] was predicted in \citep{calura2009} thanks to the delayed gas accretion in their model.}
	
	\section{Summary}
	We investigate the age-element abundance structure of the Galactic outer disc ($10<r<15$ kpc) as observed by the APOGEE survey. We construct a chemical evolution model and propose a new scenario for the chemical enrichment history of the outer disc to match the data. The key result is that our model invokes an additional {recent gas accretion event}. 
	
	We first examine the distribution of the outer disc stars in the [$\alpha$/Fe]-[Fe/H] diagram. All stars follow an anti-correlation with decreasing [$\alpha$/Fe] with increasing [Fe/H]. 
	With recently published age measurements for a large sample of stars observed by the APOGEE survey, we further utilise the distribution of stars in the three-dimensional space [$\alpha$/Fe]-[Fe/H]-Age to better constrain the chemical enrichment history of the outer disc. We identify two sequences of the [$\alpha$/Fe]-[Fe/H] relation with different ages. Surprisingly, the younger stars with ${\rm Age}<6$ Gyr are less metal-rich than the older stars.
	
	This age-metallicity anti-correlation suggests a {rather complicated} chemical enrichment history. The young stars must have formed in a metal-poor environment that has been diluted by recent, metal-poor gas accretion. Equally surprising is that the first generation of stars that formed out of the diluted gas also have enhanced $\alpha$-abundances with $[\alpha/{\rm Fe}]\sim 0.14\;$dex. This further implies that a star burst was triggered along with the gas accretion event boosting the $\alpha$-abundance through the enrichment from short-lived Type-II supernova. Moreover, the relatively low metallicities observed for this population suggest that the dilution has been very effective. This in turn implies that a relatively large amount of metal-poor gas has been accreted. We estimate the mass of the accreted gas to be roughly three times the gas reservoir in the Galactic outer disc at the epoch of accretion. 
	
	We use a full chemical evolution model to constrain this {late-accretion scenario}. We explore traditional models assuming single exponentially declining gas accretion over cosmic time {referred to as single-$\tau$ model as well as} alternative models that include an additional {short-term} gas accretion event at a cosmic time of {$8.2\;$Gyr ($z\sim0.6$) on top of the smooth, long-term accretion. The delay time is constrained by the age of the metal-rich stars in the low-$\alpha$ sequence.} We call this alternative model `{late-accretion} model'. By comparing the prediction of these models with observations in the three-dimensional parameter space [$\alpha$/Fe]-[Fe/H]-Age, we find that the {late-accretion} model {including a second accretion phase with a relatively short timescale of $\sim 0.7\;$Gyr} matched the data best.
	
	We also investigate the potential effect of radial migration and find that radial migration {helps to relieve the tension between the {single-$\tau$} model and the observational data, but is still not able to fully explain the observations.}
	{Our simulations suggest that the Milky Way accreted roughly three times the gas reservoir of the outer disc at {$z\sim 0.6$} during this late-accretion episode.} Our simulations further imply that a star burst was triggered during accretion. These characteristics imply {that the late-accretion event may have been triggered by a galaxy merger event}. If true, the data can be used to infer the properties of the accreted galaxy. The gas accretion from the infalling system must be lower than $\sim-0.5\;$dex, suggesting a dwarf galaxy with stellar mass $M_*<10^9{M_{\odot}}$ according to the mass-metallicity relation at intermediate redshift. Assuming the accreted gas distributed homogeneously within the outer disc, we estimate that the gas fraction of the accreted galaxy must have been $\sim 75$ per cent. 
	It will be interesting in future work to further explore the kinematic and dynamical effects of our {late-accretion} model on the evolution of the outer Galactic disc.  
	
	
	
	\section*{Acknowledgements}
	{We thank the anonymous referee for constructive comments and suggestions which significantly improved the clarity and robustness of the paper.}
	J.L. is grateful to Gail Zasowski for help in the introduction of the APOGEE project and useful general comments and suggestions {and Sten Hasselquist for useful discussions. } 
	
	The Science, Technology and Facilities Council is acknowledged for support through the Consolidated Grant ‘Cosmology and Astrophysics at Portsmouth’, ST/N000668/1. Numerical computations were done on the Sciama High Performance Compute (HPC) cluster which is supported by the ICG, SEPnet and the University of Portsmouth. {OZ and DAGH acknowledge support from the State Research Agency (AEI) of the Spanish Ministry of Science, Innovation and Universities (MCIU) and the European Regional Development Fund (FEDER) under grant AYA2017-88254-P.} J.G.F-T is supported by FONDECYT No. 3180210 and Becas Iberoam\'erica Investigador 2019, Banco Santander Chile.
	
	Funding for the Sloan Digital Sky Survey IV has been provided by the Alfred P. Sloan Foundation, the U.S. Department of Energy Office of Science, and the Participating Institutions. SDSS acknowledges support and resources from the Center for High-Performance Computing at the University of Utah. The SDSS web site is www.sdss.org.
	
	SDSS is managed by the Astrophysical Research Consortium for the Participating Institutions of the SDSS Collaboration including the Brazilian Participation Group, the Carnegie Institution for Science, Carnegie Mellon University, the Chilean Participation Group, the French Participation Group, Harvard-Smithsonian Center for Astrophysics, Instituto de Astrofísica de Canarias, The Johns Hopkins University, Kavli Institute for the Physics and Mathematics of the Universe (IPMU) / University of Tokyo, the Korean Participation Group, Lawrence Berkeley National Laboratory, Leibniz Institut für Astrophysik Potsdam (AIP), Max-Planck-Institut für Astronomie (MPIA Heidelberg), Max-Planck-Institut für Astrophysik (MPA Garching), Max-Planck-Institut für Extraterrestrische Physik (MPE), National Astronomical Observatories of China, New Mexico State University, New York University, University of Notre Dame, Observatório Nacional / MCTI, The Ohio State University, Pennsylvania State University, Shanghai Astronomical Observatory, United Kingdom Participation Group, Universidad Nacional Autónoma de México, University of Arizona, University of Colorado Boulder, University of Oxford, University of Portsmouth, University of Utah, University of Virginia, University of Washington, University of Wisconsin, Vanderbilt University, and Yale University.

\end{document}